\documentclass[reprint,superscriptaddress,aps,prl]{revtex4-1}
\usepackage{amsmath,epsfig,amssymb,subfigure,bm,dsfont}
\usepackage{lipsum} 
\usepackage{color}
\usepackage{graphicx}
\usepackage{dcolumn}
\usepackage{bm}
\usepackage{bbold}
\usepackage{upgreek}
\usepackage{amsfonts,color}
\usepackage{amsmath}
\usepackage[unicode=true, breaklinks=false, pdfborder={0 0 1}, backref=false, colorlinks=true, linkcolor=blue, urlcolor=blue, citecolor=blue]{hyperref}

\newcounter{lastnote}

\begin{document}
\title{Direct Observation of Antimagnons with Inverted Dispersion}

\author{Hanchen Wang}
\email{hanchen.wang@mat.ethz.ch}
\thanks{These authors contributed equally to this work.}
\affiliation{%
Department of Materials, ETH Zurich, Zurich 8093, Switzerland
}%

\author{Junfeng Hu}
\thanks{These authors contributed equally to this work.}
\affiliation{%
International Quantum Academy, Shenzhen 518048, China
}%

\author{Wenjie Song}
\thanks{These authors contributed equally to this work.}
\affiliation{%
International Quantum Academy, Shenzhen 518048, China
}%

\author{Artim L. Bassant}
\thanks{These authors contributed equally to this work.}
\affiliation{%
Institute for Theoretical Physics, Utrecht University, Utrecht 3584CC, The Netherlands
}%

\author{Jinlong~Wang}
\affiliation{%
	International Quantum Academy, Shenzhen 518048, China
}%
\affiliation{%
	Fert Beijing Institute, State Key Laboratory of Spintronics, School of Integrated Circuit Science and Engineering, Beihang University, Beijing 100191, China
}%

\author{Haishen~Peng}
\affiliation{%
International Quantum Academy, Shenzhen 518048, China
}%

\author{Emir Karad\v{z}a}
\affiliation{%
Department of Materials, ETH Zurich, Zurich 8093, Switzerland
}%

\author{Paul No\"el}
\affiliation{%
Universit\'e de Strasbourg, UMR 7504, CNRS, Institut de Physique et Chimie des Mat\'eriaux de Strasbourg, Strasbourg, 67000, France
}%

\author{William~Legrand}
\affiliation{%
Universit\'e Grenoble Alpes, CNRS, Institut N\'eel, Grenoble 38042, France
}%

\author{Richard~Schlitz}
\affiliation{%
Department of Physics, University of Konstanz, Konstanz 78457, Germany
}%

\author{Jilei~Chen}
\affiliation{%
	International Quantum Academy, Shenzhen 518048, China
}%

\author{Song~Liu}
\affiliation{%
	International Quantum Academy, Shenzhen 518048, China
}%

\author{Dapeng~Yu}
\affiliation{%
	International Quantum Academy, Shenzhen 518048, China
}%

\author{Jean-Philippe~Ansermet}
\affiliation{%
	Institute of Physics, École Polytechnique Fédérale de Lausanne (EPFL), 1015 Lausanne, Switzerland
}%
\affiliation{%
	International Quantum Academy, Shenzhen 518048, China
}%

\author{Rembert A. Duine}
\email{r.a.duine@uu.nl}
\affiliation{%
Institute for Theoretical Physics, Utrecht University, Utrecht 3584CC, The Netherlands
}%
\affiliation{%
Department of Applied Physics, Eindhoven University of Technology,
 P.O. Box 513, 5600 MB Eindhoven, The Netherlands
}%

\author{Pietro Gambardella}
\email{pietro.gambardella@mat.ethz.ch}
\affiliation{%
Department of Materials, ETH Zurich, Zurich 8093, Switzerland
}%

\author{Haiming Yu}
\email{haiming.yu@buaa.edu.cn}
\affiliation{%
Fert Beijing Institute, State Key Laboratory of Spintronics, School of Integrated Circuit Science and Engineering, Beihang University, Beijing 100191, China
}%
\affiliation{%
International Quantum Academy, Shenzhen 518048, China
}%
\date{\today}

\begin{abstract} 
We report direct spectroscopic evidence of antimagnons, i.e., negative-energy spin waves identified by their signature inverted dispersion and left-handed precession with Brillouin light scattering (BLS) spectroscopy. We investigate an ultrathin BiYIG film with a perpendicular magnetized anisotropy that compensates the demagnetizing field. By injecting a spin-orbit torque, the magnetization is driven into auto-oscillation and eventually into a non-equilibrium reversed state above a secondary current threshold ($\sim$1.2$\times$10$^7$~A/cm$^2$). The dispersion is measured by wavevector-resolved BLS and exhibits a sharp change from an upward dispersion to a downward one, in agreement with theoretical predictions and micromagnetic simulations. Around the threshold current, we observe the coexistence of conventional magnons and antimagnons. Our work establishes antimagnons with inverted dispersion and is a first step towards exploring novel phenomena and applications due to magnon-antimagnon coupling, such as magnon amplification and magnon-antimagnon entanglement, which are part of the emerging field of antimagnonics.

\end{abstract}

\maketitle

Magnons, the bosonic quasi-particles corresponding to spin-wave excitations, can transfer spin information in magnetic insulators without charge transport and are therefore promising for next-generation spintronic devices with low-power consumption~\cite{Kruglyak2010,Chumak2015,Dieny2020,Pirro2021,Yuan2022}. Conventionally, magnons are coherently excited at resonance using microwave antennas that generate large-amplitude spin waves~\cite{Vla2008,Dem2009,Yu2014}. More recently, pure spin currents injected from adjacent heavy metals have been employed to excite magnons and investigate spin transport in magnetic insulators~\cite{Cor2015,Wim2019}, as well as to drive magnetization dynamics via spin–orbit torques (SOTs)~\cite{Man2019,Dem2020}. Spin current injection occurs via the spin Hall effect of the heavy metal, which converts a charge current into a transverse spin current~\cite{sinova2015she}.
When the SOT generated by absorption of the spin current from the heavy metal overcomes the intrinsic damping~\cite{Ber1996} of the magnetic insulator, the system enters a regime of auto-oscillation, where sustained magnon excitation occurs without requiring a microwave driving field~\cite{schlitz2025autoosc,collet2016coherent,Dem2016,Aker2016,Ilya2025}. 
Ideally, the magnon intensity can be further enhanced by driving a larger current, but is often constrained by dynamic instabilities~\cite{Suhl1957,Rezende1990,Yem2025} due to nonlinear magnon scattering~\cite{Moh2021,Sheng2023,Hac2024}. 
Recent developments in the growth of magnetic insulators~\cite{Mer2024,William2025,Hanchen2025} have provided a crucial solution to overcome this obstacle by engineering perpendicular magnetic anisotropy (PMA)~\cite{Sou2018} that compensates the shape anisotropy induced by the demagnetizing field and enables large-angle circular precession~\cite{Div2019}. When the PMA compensates the demagnetizing field, a negative damping torque may drive the magnetic system beyond equilibrium and eventually towards a dynamically stabilized reversed magnetic state, as predicted by recent theoretical studies~\cite{Har2021,Ulrichs2020}.
Pioneering experiments observed signatures of inverted magnetic states in confined spin-valve and nano-oscillator systems~\cite{Kent2003,mohseni2013droplet} driven by spin-transfer torque~\cite{Ber1996,Slon1996}.
Very recently, SOT-driven dynamically stabilized inverted states have been experimentally demonstrated in extended magnetic layers using the magneto-optic Kerr effect in a magnetic insulator~\cite{karadza2026antimagnon} and spin-torque ferromagnetic resonance in a thin metal film~\cite{kurebayashi2026stability}.

In the dynamically stabilized inverted state, the magnetization is aligned against the external magnetic field. Small perturbations bring the magnetization closer to its equilibrium orientation and therefore lower the magnetic energy. These negative-energy excitations are referred to as antimagnons~\cite{Har2024,Ado2024,Yaroslav2025,liu2025antimagnon}. Unlike relativistic particle-antiparticle pairs, however, magnons and antimagnons in a ferromagnet are not related by a fundamental symmetry of a Lorentz-invariant theory; Here, the term ``antimagnon'' specifically denotes the negative-energy excitation defined relative to the dynamically stabilized inverted state. The dispersion of antimagnons is predicted to be distinct from that of conventional magnons, featuring an inverted curvature with decreasing frequency for increasing wavevectors~\cite{Har2021,Har2024}, which is the signature of antimagnons in ferromagnets as opposed to conventional magnons. Antimagnons provide unique opportunities to explore novel effects rooted in magnon-antimagnon coupling, such as magnon amplification due to the super-radiance~\cite{Har2022,Artim2024,wang2024supermirrors,errani2025negative} and magnon entanglement~\cite{Ado2024,Artim2024,Yaroslav2025}. In addition, antimagnons can provide fundamental insight into exotic phenomena such as magnonic Hawking radiation~\cite{Rol2017,Doo2019,Ado2024,Nak2024}. Despite recent theoretical and experimental work, spectroscopic evidence for antimagnons with inverted dispersion remains hitherto elusive.
\begin{figure}
\includegraphics[width=85mm]{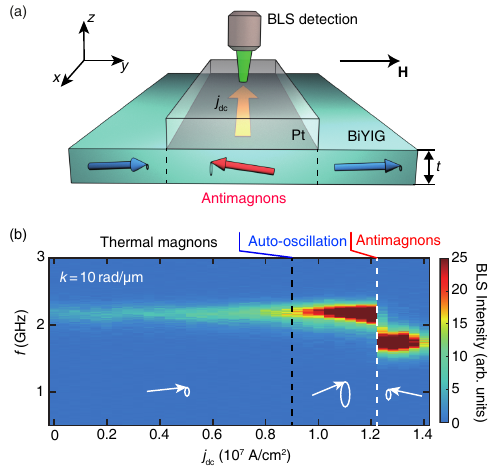}
\caption{(a) Schematic illustration of an SOT device based on a BiYIG thin film with a Pt bar on top. The BiYIG film thickness $t=4$~nm. The Pt bar is about 10~$\mu$m wide. A magnetic field of 60~mT is applied along the $y$ direction. An electric direct current (dc) current $j_{\rm dc}$ is applied along the Pt bar and exerts spin-orbit torques on BiYIG underneath to drive its magnetization into auto-oscillation and eventually into a dynamically stabilized non-equilibrium state where antimagnons emerge (red arrow). The Brillouin light scattering (BLS) technique with wavevector ($k$) resolution is employed to detect magnons. The detection laser spot is placed on the Pt/BiYIG region. (b) BLS spectra measured as a function of current density $j_{\text{dc}}$. The detection wavevector is fixed at $k=10~\text{rad}/\upmu$m. Three regimes are observed with increasing current density.}
\label{fig1}
\end{figure}

In this Letter, we report direct observation of antimagnons in a magnetic insulator with inverted dispersion driven by SOT. In contrast to spin-torque-driven bullets, which are strongly nonlinear and spatially self-localized modes below the linear spin-wave spectrum~\cite{Rip2004,Slavin2005,Bonetti2010,Jungfleisch2016,Div2017}, the excitations investigated here form a plane-wave-like, wavevector-resolved spectrum within an extended inverted magnetic state. The inverted dispersion predicted for the high-current homogeneous switched state in Ref.~\cite{Ulrichs2020}, rather than for the bullet regime itself, is closely related to the antimagnon dispersion measured here. The key ingredient for the formation of antimagnons is a compensated magnetic system. We fabricated a high-quality $4$-nm-thick BiYIG film grown on a (111)-oriented yttrium scandium gallium garnet substrate via high-temperature rf magnetron sputtering~\cite{William2025,Hanchen2025} [see Supplemental Material (SM)~\cite{SI} for details].
The PMA is engineered by tuning the tensile strain from the substrate to almost compensate the demagnetization field (see SM for ferromagnetic resonance characterization~\cite{SI}). Subsequently, a 5~nm-thick Pt layer was deposited with dc magnetron sputtering on BiYIG and patterned by e-beam lithography into a 10-$\upmu$m-wide (along $y$) and 200-$\upmu$m-long (along $x$) bar for SOT injection, as shown in Fig.~\ref{fig1}(a) and SM~\cite{SI}.
An electrical dc current of density $j_{\rm dc}$ is applied to the Pt bar (in the $-x$ direction) to exert the SOT on the underlying BiYIG. The antidamping vs damping-like character of the SOT is determined by the relative alignment of the injected spins (polarized along $\pm y$, depending on the sign of $j_{\rm dc}$) and magnetization. The latter is controlled by an external magnetic field applied along the $y$ direction.

\begin{figure*}
	\includegraphics[width=170mm]{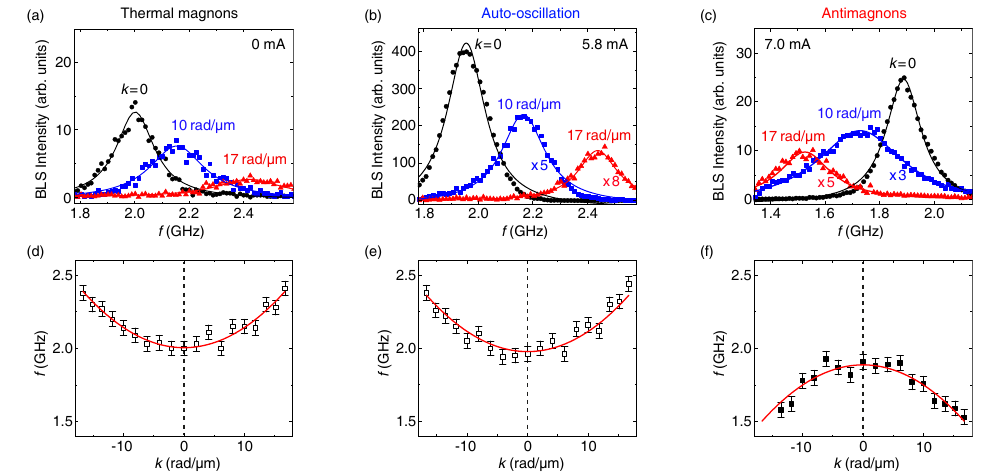}
	\caption{BLS intensity spectra measured for three different wavevectors (0, 10 and 17~rad/$\upmu$m) with (a) zero current, corresponding to thermal magnons, (b) 5.8~mA, corresponding to the auto-oscillation regime, and (c) 7.0~mA, corresponding to the antimagnon regime, where the resonance frequency exhibits a red shift at higher $k$. The intensity of the high-$k$ spectra in (b) and (c) are amplified for optimized presentation. (d-f) Magnon dispersion relations extracted from a series of BLS spectra in three characteristic regimes of (d) thermal magnons, (e) auto-oscillation and (f) antimagnons. Red solid lines represent the dispersion relations theoretically calculated for the respective cases based on Eq.~(\ref{dispersion}). All data are measured under an applied field of 60~mT.}
	\label{fig2}
\end{figure*}
We employ wavevector-resolved Brillouin light scattering (BLS) spectroscopy~\cite{Demo2001} to study the magnon dispersion of BiYIG capped by Pt. By varying the laser incident angle $\theta$ within the $xz$ plane, one can tune the magnitude of the BLS detection wavevector $k$ in the $x$ direction, e.g., $k=10$~rad/$\upmu$m for $\theta=25^\circ$. Figure~\ref{fig1}(b) shows the BLS spectra measured as a function of positive electric current density $j_{\rm dc}$ at $k=10$~rad/$\upmu$m. For a positive current, the SOT has an antidamping character. In the low-current regime, thermal magnons are detected with relatively weak BLS intensity representing the quasi-equilibrium state of the system~\cite{Cor2015}. As the current density goes above a threshold of about 0.9$\times$10$^7$~A/cm$^2$ [black dashed line in Fig.~\ref{fig1}(b)], the BLS intensity rises sharply indicating that the SOT overcomes the intrinsic damping of the system and thus drives the magnetization into auto-oscillation~\cite{Dem2012,Duan2014,Col2016,Wim2019,Kum2024}. It is noteworthy that auto-oscillation does not occur with negative current even up to -1.4$\times$10$^7$~A/cm$^2$, consistently with the inverted sign of the spin current resulting in a dampinglike torque (see SM for the full spectra~\cite{SI}). 

In conventional in-plane magnetized thin films, the dynamic demagnetizating field compresses the precession trajectory to an elliptical shape, yielding nonlinear magnon scattering that prevents further growth of the precession angle~\cite{Moh2021,Sheng2023,Hac2024}. In our PMA-compensated BiYIG film, the auto-oscillation can achieve large-angle circular precession~\cite{Mer2024,karadza2026antimagnon}. Interestingly, when $j_{\rm dc}$ further increases over a secondary critical value of $\sim$1.2$\times$10$^7$~A/cm$^2$ [white dashed line in Fig.~\ref{fig1}(a)], an entirely new phase emerges exhibiting an abrupt drop of resonance frequency from approximately 2.2~GHz to 1.7~GHz. This phase corresponds to an inverted magnetization state dynamically stabilized by negative damping~\cite{Har2021,Har2024,karadza2026antimagnon,kurebayashi2026stability}. By driving even more current in Pt, we find that at $\sim$1.4$\times$10$^7$~A/cm$^2$ the BLS intensity starts to decrease rather quickly, down to almost the same level of thermal magnons (see SM for the intensity plots~\cite{SI}). This decrease indicates that, once the magnetization is reversed, further enhancement of the antidamping SOT drives the magnetic system toward increasingly negative temperatures with more spins in the higher-energy antialigned state than in the lower-energy aligned state.

The collective spin excitations in the inverted magnetic state are called antimagnons~\cite{Har2021,Har2024} and are predicted to exhibit a signature inverted dispersion, in stark contrast with conventional magnons. In order to probe the evolution of the magnon dispersion across the dynamical stabilization threshold, we performed wavevector ($k$)-resolved BLS measurements at three characteristic current values of  $I_{\rm dc}=0.0$, 5.8, and 7.0~mA (see SM for BLS methods~\cite{SI}). Figure~\ref{fig2}(a) shows BLS intensity spectra of thermal magnons measured at three different wavevectors of $k=0$ ($\theta=0$), 10~rad/$\upmu$m ($\theta=25^\circ$) and 17~rad/$\upmu$m ($\theta=45^\circ$) under an external magnetic field of 60~mT. At higher $k$, the resonance frequency shifts up as expected for thermal magnons. More BLS spectra at several different wavevectors are shown in the SM~\cite{SI}, from which one can plot the resonance frequency $f$ as a function of $k$ in Fig.~\ref{fig2}(d) with $+k$ data extracted from anti-Stokes and $-k$ data from Stokes peaks. The dispersion is fitted with an exchange stiffness $A=2.8$~pJ/m, close to the values reported in previous works~\cite{Beach2023}. When the current is raised to 5.8~mA [Fig.~\ref{fig2}(b)], the system enters auto-oscillation with strongly enhanced BLS intensity. At higher $k$, the frequency shows again a blue shift. The corresponding dispersion [Fig.~\ref{fig2}(e)] remains conventional and can be fitted with an identical curve as that of thermal magnons in Fig.~\ref{fig2}(d). Strikingly, when the current rises to 7.0~mA [Fig.~\ref{fig2}(c)], the excitation frequency decreases at higher $k$, i.e., it shows an anomalous redshift instead of the conventional magnon blueshift reported in Figs.~\ref{fig2}(a) and (b). To quantify the dispersion, we extract the peak BLS intensity by performing Lorentz fits to a series of BLS spectra (see SM~\cite{SI}) and plot the resonance frequency $f$ as a function of the wavevector $k$. In Fig.~\ref{fig2}(f), we clearly observe an inverted dispersion curve which is a hallmark for antimagnons according to theoretical predictions~\cite{Har2021,Har2024}. 
\begin{figure}[t]
\includegraphics[width=85mm]{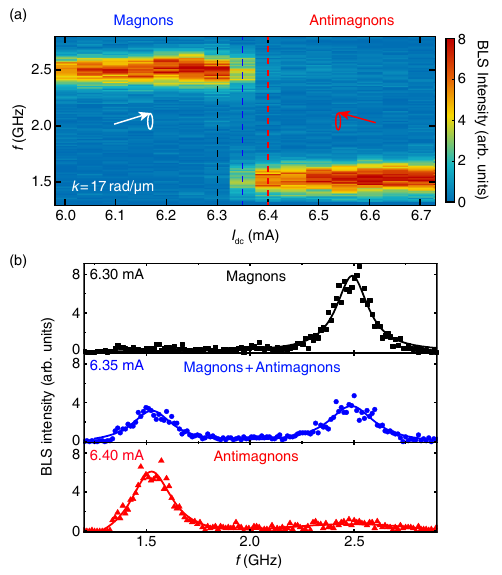}
\caption{(a) BLS spectra measured at $k=17$~rad/$\upmu$m while sweeping the current $I_{\rm{dc}}$ from 6.0 to 6.7~mA in increments of 0.05~mA. The external magnetic field is 60~mT. Around 6.35~mA (blue dashed line), the switching from the conventional magnon to antimagnon state occurs accompanied by a sharp change in the resonance frequency. (b) Three representative BLS spectra plotted for $I_{\rm{dc}}=6.30$, 6.35 and 6.40~mA around the transition [black, blue and red dashed lines in (a)]. Around 6.35~mA, two peaks are observed indicating the coexistence of the magnon and antimagnon phase.}
\label{fig3}
\end{figure}
As antimagnons correspond to linear fluctuations around the non-equilibrium reversed magnetization, their dispersion relation can be derived from the Landau–Lifshitz–Gilbert (LLG) equation~\cite{Har2024}. In the limit of negligible dipolar interactions, appropriate for our 4~nm-thick BiYIG film, the dispersion relation simplifies to
\begin{equation}
f(k) = -\frac{\gamma}{2\pi} D k^2 + \Delta,
\label{dispersion}
\end{equation}
where $\gamma$ is the gyromagnetic ratio, $D=2A/M_{\rm{s}}$ with $A$ being the exchange stiffness and $M_{\rm{s}}$ the saturation magnetization, and $\Delta$ represents the frequency of the Kittel mode corresponding to ferromagnetic resonance. The experimental data in Fig.~\ref{fig2}(f) are fitted remarkably well using the antimagnon dispersion Eq.~(\ref{dispersion}) with the same exchange stiffness $A=2.8$~pJ/m and  $M_{\rm{s}}=100$~kA/m as required for magnons in Figs.~\ref{fig2}(d) and (e). The ferromagnetic resonance frequency $\Delta=1.89$~GHz is close but not identical to the one measured for magnons at zero current, which we attribute to the effect of Joule heating on the magnetization. Thus, the results shown in Fig.~\ref{fig2}(f) provide direct evidence for the emergence of antimagnons driven by SOT. Additional characterization of the dispersion under different external magnetic fields, e.g., 40~mT and 70~mT is provided in the SM~\cite{SI}.

From the measured dispersions in Figs.~\ref{fig2}(d-f), we learn that the frequency variation between magnon and antimagnon states becomes more pronounced at higher $k$. Therefore, we push our BLS detection wavevector up to 17~rad/$\upmu$m with a laser incident angle around 45$^\circ$.
We then measure the BLS spectra as a function of applied current ranging from 6.0~mA up to 6.7~mA in steps of 0.05~mA, as shown in Fig~\ref{fig3}(a). Above the threshold current $\sim$6.35~mA, a significant frequency drop of about 1~GHz is detected from $\sim$2.5~GHz to $~\sim$1.5~GHz. In Fig.~\ref{fig3}(b), we extract three characteristic BLS spectra around the current threshold. At $6.30~\mathrm{mA}$, the signal peak centers around $2.5~\mathrm{GHz}$ corresponding to the conventional magnon state, where the magnetization is just about to, but still yet to reverse. Remarkably, at the critical point ($\sim6.35~\mathrm{mA}$)~\cite{Wan2025}, two spectral peaks emerge, reflecting the coexistence of magnons and antimagnons in critical conditions. This coexistence has not been predicted by theory and is ascribed to the dynamic instability of the system in a chaotic regime~\cite{Lee2004,Ulrichs2020}. At $6.40~\mathrm{mA}$, the BLS intensity shifts almost entirely to the low-frequency level at $\sim$1.5~GHz, showing that the system has completed the transition into the antimagnon regime following complete reversal of the magnetization. %

To further understand the formation of antimagnons and their coexistence with conventional magnons, we perform micromagnetic simulations of the magnetization dynamics using MuMax$^3$~\cite{MuMax2014} taking into account a spatially uniform dampinglike spin torque~\cite{Ulrichs2014,Ulrichs2020} to model the spin-wave dispersion across the three different regimes, i.e., thermal magnons, auto-oscillation and antimagnons. The simulated spin-wave dispersions in these regimes show good qualitative agreement with the experimental results presented in Fig.~\ref{fig2}. Simulation details and these corresponding spin-wave spectra are provided in the SM~\cite{SI}. 
Upon approaching the reversal threshold, the simulated dispersion reveals the simultaneous appearance of both magnon and antimagnon branches~\cite{SI}. A spatial snapshot of the $M_{\rm y}$ magnetization component taken at a steady state dynamics indicates that the dynamic magnetization texture is spatially inhomogeneous, with some regions pointing along the external field directions and others opposite to it~\cite{SI}. These regions are populated by magnons and antimagnons, respectively. 
In addition, spatially resolved BLS measurements indicate that sample inhomogeneities can lead to variations in the critical current at different positions~\cite{SI}, which may contribute to the observed coexistence.
Notably, for sufficiently large SOT currents, the simulations also show that the magnetization dynamics re-enters a quasi-linear regime in which the magnetization precesses with a small cone angle around the direction opposite to the applied field, in agreement with the decrease of the BLS intensity observed in Fig.~\ref{fig1}(b) at $j_{\rm dc}=1.4\times 10^7$~A/cm$^2$. 

\begin{figure}[t]
\includegraphics[width=85mm]{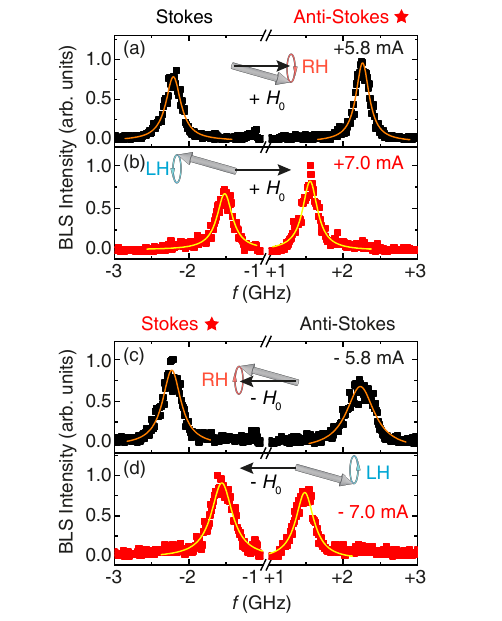}
\caption{(a) BLS spectra measured at $+60$~mT for $I_{\rm dc}=+5.8$~mA, where conventional magnons exist in the system. Due to their right-handed precession, the anti-Stokes peak is stronger than the Stokes peak. (b) BLS spectra measured at $+60$~mT for $I_{\rm dc}=+7.0$~mA, where the dynamically stabilized state is established and antimagnons are present. The anti-Stokes peak remains stronger than the Stokes peak even though the magnetization is reversed, indicating that antimagnons possess left-handed precession. (c,d) BLS spectra measured at $-60$~mT for $I_{\rm dc}=-5.8$~mA and $I_{\rm dc}=-7.0$~mA, respectively. The solid curves represent Lorentzian fits to the measured spectra. The wavevector is fixed at 17~rad/$\upmu$m.} 
\label{fig4}
\end{figure}

An important property of antimagnons is that they carry spin angular momentum opposite to that of conventional magnons in the equilibrium state. 
To experimentally verify the magnon handedness, we performed polarization-sensitive BLS measurements by analyzing the asymmetry between Stokes and anti-Stokes scattering intensities, which depends on the magnon precession chirality~\cite{Kim2020}.
As a reference, we first measured the conventional magnon branch in the auto-oscillation regime ($I_{\rm dc}=5.8$ mA), as shown in Fig.~\ref{fig4}(a). For conventional magnons in a ferromagnet, the low-energy precession is right-handed with respect to the magnetization. Therefore, magnetization reversal by the external magnetic field inverts the asymmetry of the Stokes and anti-Stokes peak intensities, as shown in Fig.~\ref{fig4}(c). 
In contrast, upon driving the system into the dynamically stabilized state at high current ($I_{\rm dc}=7.0$~mA), the asymmetry between Stokes and anti-Stokes peaks does not invert, as shown in Figs.~\ref{fig4}(b) and (d), even though the magnetization has reversed~\cite{karadza2026antimagnon,kurebayashi2026stability}. This observation indicates that excitations of the dynamically stabilized state carry opposite spin angular momentum relative to conventional magnons. This behavior is observed across different wavevectors~\cite{SI}, providing direct evidence that the observed branch has opposite handedness relative to conventional magnons, consistently with the definition of antimagnons ~\cite{Har2021,Har2024}.

In summary, we have provided direct experimental evidence for antimagnons with inverted dispersion and left-handed precession driven by SOT above a current threshold ($\sim$1.2$\times10^7$ A/cm$^2$) in a Pt/BiYIG bilayer device. By employing wavevector-resolved BLS spectroscopy, we detect the systematic evolution of magnetization dynamics while continuously raising the antidamping SOT across three successive regimes of thermal magnons, auto-oscillation and antimagnons. The dispersion relations are measured by the BLS and show an abrupt change when the system enters the antimagnonic regime, from a positive curvature for conventional magnons to a fully-inverted downward curvature for antimagnons. This experimental observation is in good agreement with theoretical predictions and is further corroborated by our micromagnetic simulations. At the largest SOT current applied, the system can be dynamically stabilized by the antidamping SOT into a quasi-linear antimagnonic regime. By tuning the SOT precisely to the critical point, we observe the coexistence of both magnons and antimagnons. Our work opens a route to explore novel phenomena related to magnon-antimagnon pair production, such as magnon amplification~\cite{Har2022} and magnon-antimagnon entanglement~\cite{Artim2024,Yaroslav2025}. SOT-driven antimagnons in a fully-compensated magnetic film allow for exploring spin-wave dynamics in non-equilibrium magnetic states beyond traditional spintronic and magnonic systems.\\

\begin{acknowledgments}
\textit{Acknowledgments}$-$ We thank I.~N. Krivorotov and J.~Xiao for helpful discussions. We wish to acknowledge the support by the NSF China under Grants No. 12525406, No. 12474104 and No. 12550004; the National Key Research and Development Program of China, Grants No. 2022YFA1402801. H.W., E.K., and P.G. acknowledge the support of the Swiss National Science Foundation (Grant No. 200021-236524). H.W. acknowledges the support of the China Scholarship Council (CSC, Grant No. 202206020091). J.H. acknowledges the support from the Shenzhen Science and Technology Program (Grant No. RCBS20231211090814026). A.L.B. and R.A.D. thank the Dutch Research Council (NWO) for funding via the projects “Black holes on a chip” with project number OCENW.KLEIN.502 and “Fluid Spintronics” with project number VI.C.182.069. R.S. acknowledges funding by the Deutsche Forschungsgemeinschaft (DFG, Grant No. 425217212). 
\end{acknowledgments}
\bibliographystyle{unsrt}

\end{document}